\documentclass[12pt]{article}
\usepackage[english]{babel}
\usepackage{amssymb}
\usepackage{amsmath}
\usepackage{amsfonts}
\usepackage{indentfirst}
\usepackage{graphicx}

\usepackage{color}

\def\e{{\rm e}}
\def\d{{\rm d}}
\def\1{{\rm 1}}
\def\d{{\rm d}}

\def\r{\mathbf{r}}
\def\bfc{\mathbf{c}}

\begin{document}
\title{Binding of muonated hydrogen molecules on the occasion of the Born-Oppenheimer approximation\\ 90th anniversary}
\author{A.J.C. Varandas\footnote{School of Physics and Physical Engineering, Qufu Normal University, 273165 Qufu, China, and Chemistry Centre and Department of Chemistry, University of Coimbra, 3004-535 Coimbra, Portugal},
J.~da Provid\^encia\footnote{Department of Physics, University of Coimbra, P 3004-516 Coimbra, Portugal (providencia@teor.fis.uc.pt)}~
and J.P. da Provid\^encia\footnote{Department of Physics, University of Beira Interior, P-6201-001 Covilh\~a, Portugal
(joaodaprovidencia @daad-alumni.de)}}
\date{(June 1, 2018)}
\maketitle
\begin{abstract}
The stability of four fermionic particles with unit charge, of which, two are positively, and two negatively charged, is discussed.
Except for using the simplest approximation of a single Gaussian orbital per particle, the problem is exactly solved variationally and, by varying the masses to simulate molecular di-hydrogen, mono-muonated di-hydrogen and di-muonated di-hydrogen, employed to illustrate the celebrated Born-Oppenheimer approximation on the occasion of its 90th anniversary. It is suggested that it is valid only for di-hydrogen.
\end{abstract}

\section{Introduction}
The understanding of Physics and Chemistry bears on a hierarchy of approximations, with the one carrying the names of Born and Oppenheimer~\cite{BOR27:457} (BO) being most celebrated. At the dusk of its ninetieth anniversary, we wish to illustrate the relevance of the BO approximation by considering a four-fermionic model, treated beyond such an approximation, although preserving its spirit.

Great interest in $\mu$-mesonic molecules arose in connection with cold fusion processes.
Muons are leptons, like electrons, but are about 200
times heavier. If an electron is replaced by a muon in a hydrogen molecule,
the nuclei become about 200 times closer \cite{cohen}, and
when the nuclei are so near to each other, the probability of nuclear fusion increases enormously.

Our aim is to understand the stability of the $\mu$-mesonic molecules on the
basis of simple models satisfying the important requirement of
consistently describing atoms and molecules of {hydrogen}. By
consistent description we mean that the schematic model wave
function of the $\mu$-mesonic molecule should contain as factors the
schematic model wave functions of the atoms, at least
for some particular value of parameters specifying correlations
between electrons and protons, or between muons and protons,
belonging to distinct atoms.

{It should be noted at this point that the idea of allocating wave-functions to nuclei as well as electrons is not new, with the reader being addressed to the literature~\cite{[8],[9],[10]}
for details. Subtleties have to be consider, namely due to the fact that some translation of the molecule is included unless special steps are taken to exclude it~\cite{[11]},
a point that will be addressed later in the subsection titled ``centre of mass correction''. A furher remark to note that the beyond Born-Oppenhemier approximation here discussed makes no attempt to perform dynamics on potential-energy surfaces.}

All calculations here reported will be performed in atomic units (a.u.): of length, $\rm a_0=0.529177\,nm$; of energy, $\rm E_h=27.211652\,eV$.
\allowdisplaybreaks{
\section{Preliminary}
The Hamiltonian of the mono-muonic hydrogen molecule reads,
\begin{eqnarray*}&&H=-{1\over2}\nabla_1^2 -{1\over2\mu}\nabla_2^2-{1\over2M}\nabla_3^2-{1\over2M}\nabla_4^2
+{1\over|\r_1-\r_2|} +{1\over|\r_3-\r_4|}\\&&-{1\over|\r_1-\r_3|}-{1\over|\r_1-\r_4|}-{1\over|\r_2-\r_3|}-{1\over|\r_2-\r_4|},
\end{eqnarray*}
where $\mu$ is the muon mass, $M$ is the proton mass, and the electron mass is 1 in atomic units, which are here used.}
We describe the mono-muonic hydrogen molecule by a two-center model assuming that the
electron orbits around one of the protons, and the muon orbits around the other proton. We further assume that the
orbitals of the electron, the proton to it associated, and of the muon and proton to it associated are, respectively,
$$\e^{-\alpha(\r-\bfc/2)^2},~\e^{-\beta(\r-\bfc/2)^2},~\e^{-\gamma(\r+\bfc/2)^2},~\e^{-\beta'(\r+\bfc/2)^2},$$
where $\bfc$ is a constant vector. The wave function we consider is the symmetrized product of these orbitals,
$$\Phi(\r_1,\r_2,\r_3,\r_4)={\cal S}~\left(\e^{-\alpha(\r_1-\bfc/2)^2}~\e^{-\gamma(\r_2+\bfc/2)^2}~\e^{-\beta(\r_3-\bfc/2)^2}~\e^{-\beta'(\r_4+\bfc/2)^2}\right),$$
where $\cal S$ is the symmetrizing operator,
$${\cal S}~\Psi(\r_1,\r_2,\r_3,\r_4)=\Psi(\r_1,\r_2,\r_3,\r_4)+\Psi(\r_2,\r_1,\r_3,\r_4)+\Psi(\r_1,\r_2,\r_4,\r_3)+\Psi(\r_2,\r_1,\r_4,\r_3).$$
and the parameters $\alpha,~\gamma,~\beta,~ \beta'$ and $c= |\bfc|$ are all variationally determined. Thence, the form and symmetry adopted for the wave-function are assumed to have enough flexibility for closely approaching maximum stability for the species under scrutiny. The relevant expectation values are expressed by integrals over the coordinates of the four point particles,
which are products of integrals involving at most two particles at a time.

If, in our model, we replace the muon mass by the electron mass, a model for the hydrogen molecule (di-hydrogen) is obtained.
Instead, if the electron mass is replaced by the muon mass, a model for the di-muonated hydrogen molecule is obtained. Moreover, if the
muon and the proton masses are replaced by the electron mass, a model for the positronium molecule \cite{wheeler,bao,varandas,varandas} is obtained.

We recall that \cite{abramowitz}
\begin{eqnarray*}
J(r')=\int\d^3\r{\e^{-\r^2}\over\sqrt{(\r-\r')^2}}=\pi^{3/2} {{\rm erf}(r')\over r'}
.\end{eqnarray*}

\section{Rayleigh quotients}
In order to compute its expectation value, it is convenient to express
the  Hamiltonian $H$ as a sum of terms,
$$H=K_{ee}+K_{PP}+V_{ee}+V_{PP}+V_{eP},$$
where
\begin{eqnarray*}&&K_{ee}=-{1\over2}\nabla_1^2 -{1\over2\mu}\nabla_2^2,\quad K_{PP}=-{1\over2M}\nabla_3^2-{1\over2M}\nabla_4^2,\\
&&V_{ee}={1\over|\r_1-\r_2|},\quad V_{PP}= {1\over|\r_3-\r_4|},
\\&&V_{eP}=-{1\over|\r_1-\r_3|}-{1\over|\r_1-\r_4|}-{1\over|\r_2-\r_3|}-{1\over|\r_2-\r_4|}.
\end{eqnarray*}
Here, the index $e$ stands for {lepton} and the index $P$ for proton, so that,
for instance, $K_{ee}$ is the kinetic energy of the two {lepton} (electron, muon) subsystem
and $K_{PP}$ is the kinetic energy of the two proton subsystem. In the following, the relevant
Rayleigh {\color{red}quotients} are expressed in terms of the integrals listed in the Appendix.
We obtain
\begin{eqnarray*}
{\langle\Phi| K_{ee}|\Phi\rangle\over\langle\Phi |\Phi\rangle }&=&{{1\over2}I^{ee}_{KD}(\alpha,\gamma)+{1\over2\mu}I^{ee}_{KD}(\gamma,\alpha)
+({1\over2}+{1\over2\mu})I^{ee}_{KE}(\alpha,\gamma)\over I^{ee}_{ND}(\alpha,\gamma)+ I^{ee}_{DE}(\alpha,\gamma)},\\
{\langle\Phi| K_{PP}|\Phi\rangle\over\langle\Phi |\Phi\rangle }&=&{1\over2 M}~{I^{PP}_{KD}(\beta,\beta')+I^{PP}_{KD}(\beta',\beta)
+2I^{PP}_{KE}(\beta,\beta')\over I^{PP}_{ND}(\beta,\beta')+I^{PP}_{NE}(\beta,\beta')},\\
{\langle\Phi| V_{ee}|\Phi\rangle\over\langle\Phi |\Phi\rangle }&=&
{I^{ee}_{VD}(\alpha,\gamma)+I^{ee}_{VE}(\alpha,\gamma)\over I^{ee}_{ND}(\alpha,\gamma)+I^{ee}_{NE}(\alpha,\gamma)},\\
{\langle\Phi| V_{PP}|\Phi\rangle\over\langle\Phi |\Phi\rangle }&=&
{I^{PP}_{VD}(\beta,\beta')+I^{PP}_{VE}(\beta,\beta')\over I^{PP}_{ND}(\beta,\beta')+I^{PP}_{NE}(\beta,\beta')},\\
%
{\langle\Phi| V_{eP}|\Phi\rangle\over\langle\Phi |\Phi\rangle }&=&-{{\cal N}_{eP}\over {\cal D}_{eP}},.
\end{eqnarray*}
where
\begin{eqnarray*}&&
{\cal N}_{eP}=
I^{ePD}_{VDD}(\alpha,\beta')I^{eP}_{NDD}(\gamma,\beta)
+I^{ePD}_{VDD}(\gamma,\beta)I^{eP}_{NDD}(\alpha,\beta')
\\&&
+I^{ePS}_{VDD}(\alpha,\beta)I^{eP}_{NDD}(\gamma,\beta')
+I^{ePS}_{VDD}(\gamma,\beta')I^{eP}_{NDD}(\alpha,\beta)
\\&&
+2I^{eP}_{VDE}(\alpha,\beta,\beta')I^{eP}_{NDE}(\gamma,\beta',\beta)+2I^{eP}_{VDE}(\gamma,\beta',\beta)I^{eP}_{NDE}(\alpha,\beta,\beta')\\&&
+2I^{eP}_{VED}(\alpha,\gamma,\beta)I^{eP}_{NED}(\gamma,\alpha,\beta')+2I^{eP}_{VED}(\gamma,\alpha,\beta')I^{eP}_{NED}(\alpha,\gamma,\beta)\\&&
+4I^{eP}_{VEE}(\alpha,\gamma,\beta,\beta')I^{eP}_{NEE}(\alpha,\gamma,\beta,\beta'),
\end{eqnarray*}
and
\begin{eqnarray*}&&
{\cal D}_{eP}=
I^{eP}_{NDD}(\alpha,\beta)I^{eP}_{NDD}(\gamma,\beta')
+I^{eP}_{NDE}(\alpha,\beta,\beta')I^{eP}_{NDE}(\gamma,\beta',\beta)
\\&&
+I^{eP}_{NED}(\alpha,\gamma,\beta)I^{eP}_{NED}(\gamma,\alpha,\beta')
+I^{eP}_{VEE}(\alpha,\gamma,\beta,\beta')I^{eP}_{NEE}(\gamma,\alpha,\beta',\beta)
.\end{eqnarray*}
\section{Results}
In order to describe two interacting hydrogen atoms, we replace, in the Hamiltonian, the muon mass $\mu$ by 1, and
in order to describe two interacting di-muonated hydrogen atoms, we replace the electron mass, equal to 1, by
the muon mass $\mu$. Following a procedure similar to the one described elsewhere~\cite{varandas,varandas2}, we obtain in the former case:
\begin{itemize}
\item Binding energy of two hydrogen atoms: $-0.125949$ a.u.;

\item If $1.626<c<1.627$, the energy of the hydrogen molecule is: $-0.936496$ a.u.;

\item If $c\rightarrow\infty,$ the energy of the hydrogen molecule approaches $-0.810547$ a.u.;

\item Distance between the centres in the hydrogen molecule: 1.626 a.u.;

\item Radius of the electron orbit in the atoms of the hydrogen molecule: 1.50 a.u.;

\item Radius of the proton orbit in the atoms of the hydrogen molecule: 0.229 a.u.;
\end{itemize}
The results so obtained are graphically summarized in Figure~1. 

\begin{figure}[h]
\centering
\includegraphics[width=.5\textwidth, height=0.3\textwidth]
{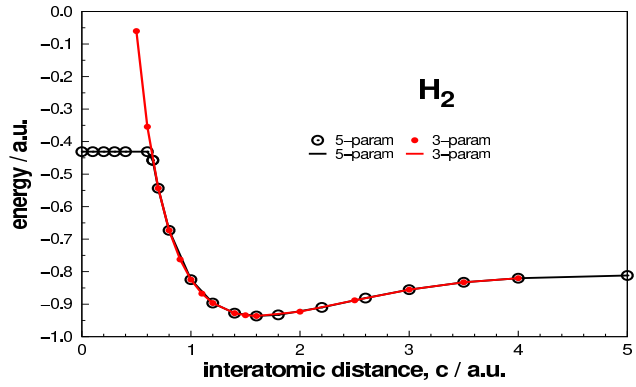}
\caption{\label{fig1}Di-hydrogen (hydrogen) molecule: energy vs bond distance.
The attraction between the atomic centers shows up for $c>0.5$. atomic units are used.}
\end{figure}


Similarly, for two interacting muonic hydrogen atoms, we replace in the Hamiltonian the electron rest mass ($\rm 1\,a.u.$)  by the muon mass $\mu$, yielding:
\begin{itemize}
\item Binding energy of two muonic hydrogen atoms: $-7.6337$ a.u.;

\item If $0.0142<c<0.0143$, the energy of the di-muonic hydrogen molecule is: $-106.059$ a.u.;

\item If $c\rightarrow\infty,$ the energy of the di-muonic hydrogen molecule approaches $-98.4153$ a.u.;

\item Distance between the centres in the muonic hydrogen molecule: 0.0142 a.u.;

\item Radius of the muon orbit in the atoms of the muonic hydrogen molecule: 0.0112 a.u.;

\item Radius of the proton orbit in the atoms of the muonic hydrogen molecule: 0.0065 a.u.;
\end{itemize}
with the results shown in Figure~2. 

In the description of the hydrogen molecule ($\rm H_2$), it is natural to take $\alpha=\gamma,~\beta=\beta'.$
As summarized above, the radius of the electron orbital is predicted to be almost one order of magnitude larger than
the radius of the proton orbital it surrounds. In agreement with the
conventional BO approximation, we may then replace the orbital of that proton by a delta function.
In turn, the radius of the muon orbital is only slightly larger than the radius of the proton orbital it surrounds, and hence it is not reliable to replace the
orbital of that proton by a delta function. Thence, the BO approximation is not expected to be reliable for both the mono- and di-muonated di-hydrogen molecules,
as observed  in \cite{varandas2}. Indeed, even if it might be valid for treating a single hydrogen atom it is no
longer true when combined with a muonated proton since for this the proton-to-muon mass ratio is only about 10 ~\cite{varandas}.
\begin{figure}[h]
\centering
\includegraphics[width=.5\textwidth, height=0.3\textwidth]
{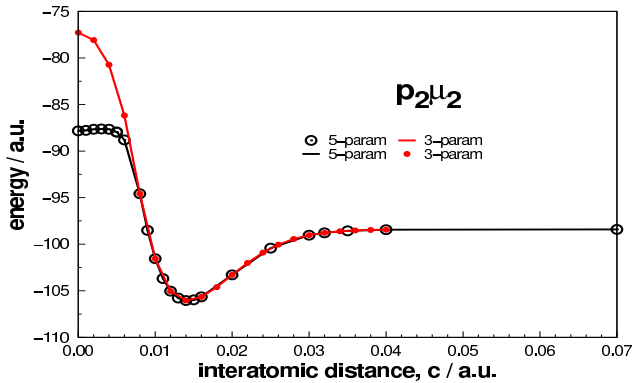}
\caption{
Di-muonated di-hydrogen (hydrogen) molecule: energy vs separation distance. The onset for attraction
between the atomic centers shows up for $c>0.006$ Symbols and units as in Fig. 1. }
\label{fig3}
\end{figure}
A chemical bond between a normal hydrogen atom and a muonated hydrogen atom does not exist because
there is no binding energy for the mono-muonated di-hydrogen molecule, as shown in Figure 3. This might be expected,
in agreement with the discussion in \cite{varandas2}, because the chemical bond is the manifestation of a resonance effect
which does not occur for the mono-muonated hydrogen.
In the cases of the di-hydrogen and hydrogen di-muonated, the four  wave-functions $\Phi(\r_1,\r_2,\r_3,\r_4), ~\Phi(\r_2,\r_1,\r_4,\r_3),~ \Phi(\r_2,\r_1,\r_3,\r_4)$ and  $\Phi(\r_1,\r_2,\r_4,\r_3)$ have the same energy. So, there exists a linear combination of them which dramatically lowers the energy and another one which raises it. Note that $\Phi(\r_1,\r_2,\r_3,\r_4)$ is equivalent to $\Phi(\r_2,\r_1,\r_4,\r_3)$, while $\Phi(\r_2,\r_1,\r_3,\r_4)$ is equivalent to $\Phi(\r_1,\r_2,\r_4,\r_3)$. In the case of the hydrogen mono-muonated, the wave-functions $\Phi(\r_1,\r_2,\r_3,\r_4),~ \Phi(\r_2,\r_1,\r_4,\r_3)$ have different anergies from the the wave functions $\Phi(2,1,3,4),~ \Phi(1,2,4,3)$  and the difference is important because the muon mass is much larger than the electron mass. So, the respective energy increase and decrease are not significant, and no resonance effect occurs.
\color{black}
Note further that to describe di-positronium, one must consider $\mu=M=1$. It is also  natural to
 take $\alpha=\gamma=\beta=\beta'$. A final remark to emphasize that if we wish to apply the conventional BO approximation we just let $\beta, \beta'\rightarrow\infty.$ It then appears that the present calculations go beyond the BO approximation, which is so unless the orbital of the heavy particles are constrained to delta functions. Indeed, in the conventional BO approximation, the protons are kept fixed while in the present calculations
we keep fixed the atomic centers around which positively and negatively charged particles describe their orbitals.

\begin{figure}[h]
\centering
\includegraphics[width=.5\textwidth, height=0.3\textwidth]
{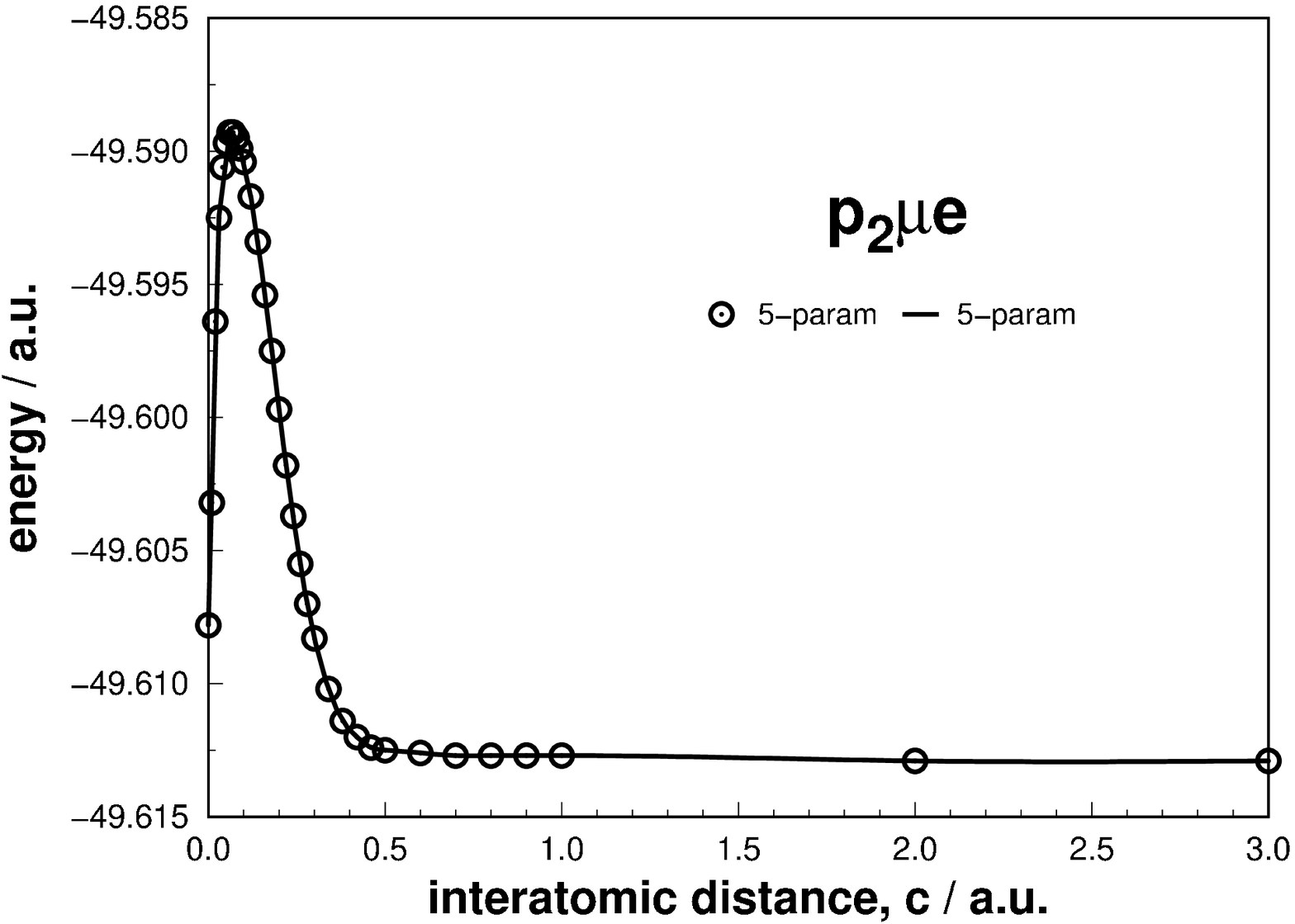}
\caption{
The energy of mono-muonic molecule without CM corrections. For $c=0,$ the energy is $-49.6078$ a.u.. Fot $c=\infty,$ the energy is $-49.6129$ a.u.. }
\label{fig4}
\end{figure}

Regarding Figures 1 and 3, the following considerations are in order. In the description of
$\rm H_2$ and the di-muonated di-hydrogen molecules, one expects that $\alpha=\gamma$
and $\beta=\beta'$. Indeed this is what happens
if the distance between the atomic centers is not too small.
However, for  small distance between the atomic centers, the energy is lowered for
$\alpha\neq\gamma$ and $\beta\neq\beta'$.
By breaking a symmetry, effective correlations
are introduced which, very often,  lower the energy \cite{peierls}.
It may also be noted that, according to the famous Gell-Mann and Brueckner theory of the high density electron gas \cite{gell-mann},
further important correlations must be considered at very small interatomic distances.
Thus, significant quantitative meaning cannot be attributed to the curves in Figures 1, 2 and 3 for
values of $c<0.5,$ $c<0.005$ and $c<0.06,$ or so, respectively for di-hydrogen, di-muonated hydrogen and mono-muonated hydrogen, on account of the
strong  correlations,
leading to the screening of the interaction,  which arise there and were not taken into account in the present work,
\color{black}

Since the equilibrium bond distance is very much reduced in the di-muonated hydrogen
molecule, the corresponding moment of inertia is accordingly much reduced.
Thus, such a molecule can hardly be expected to exhibit a rotational spectrum, because
this would not be supported by the binding potential.


For non-muonic hydrogen and di-muonic hydrogen, the spin component of the wave function is antisymmetric,
both for leptons and protons, and hence is given by
$$s_\uparrow(\sigma_1)s_\downarrow(\sigma_2)- s_\uparrow(\sigma_2)s_\downarrow(\sigma_1)$$
for the leptons, and by
$$s_\uparrow(\sigma_3)s_\downarrow(\sigma_4)-s_\uparrow(\sigma_4)s_\downarrow(\sigma_3)$$
for the protons.
\color{black} Here, by $s_\uparrow(\sigma)$ and by $s_\downarrow(\sigma)$, with $\sigma=\pm1,$ we denote the
spin eigenvectors for spin up and spin down states, respectively.
\color{black}
For mono-muonic hydrogen, the spin components of the wave function is antisymmetric
only for protons, and hence is given by any one of the following possibilities
$$s_\uparrow(\sigma_1)s_\uparrow(\sigma_2), \quad s_\downarrow(\sigma_2)s_\downarrow(\sigma_1),\quad s_\uparrow(\sigma_1)s_\downarrow(\sigma_2), \quad-s_\uparrow(\sigma_2)s_\downarrow(\sigma_1),
$$
for the leptons, and by
$$s_\uparrow(\sigma_3)s_\downarrow(\sigma_4)-s_\uparrow(\sigma_4)s_\downarrow(\sigma_3)$$
for the protons.

\begin{figure}[ht]
\centering
\includegraphics[width=.5\textwidth, height=0.3\textwidth]
{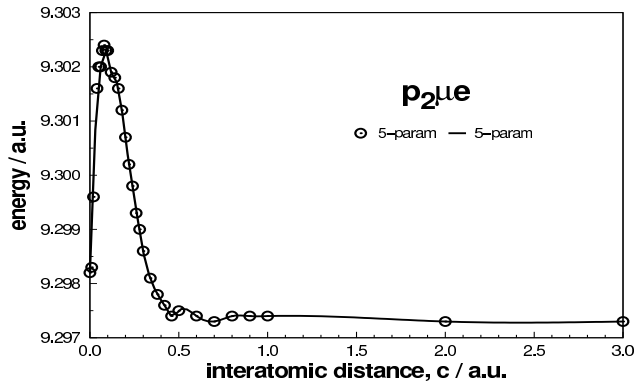}
\caption{
The center of mass energy of the mono-muonic molecule. }
\label{fig6}
\end{figure}

\begin{figure}[h]
\centering
\includegraphics[width=.5\textwidth, height=0.3\textwidth]
{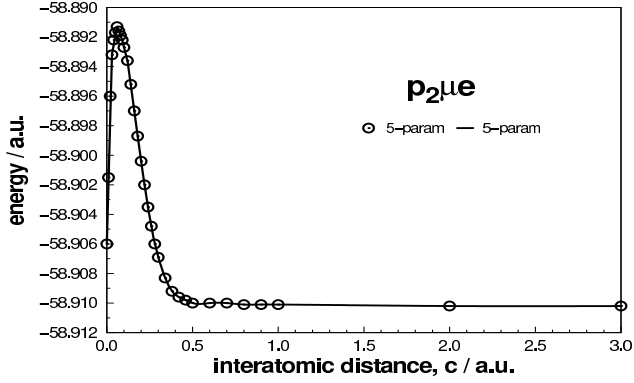}
\caption{
The energy of mono-muonic molecule with CM correction. For $c=0,$ the energy is$-59.906$ a.u.. For $c=infty,$ the energy is $-58.9106$ a.u..}
\label{fig5}
\end{figure}


\section*{Center of mass correction}
\def\bfp{{\bf p}}
\def\bfr{{\bf r}}
The Hamiltonian in Eq.~(1) has translation symmetry, that is, it commutes with the total momentum,
$${\bf P}=-i\nabla_1-i\nabla_2\mu-i\nabla_3M-i\nabla_4M.$$ Recall that the mass of the electron is $\rm 1\,a.u.$
Then, we may define the internal (intrinsic) Hamiltonian as
$$H_{int}=H-\frac{{\bf P}^2}{2M_{tot}},$$
where $M_{tot}=1+\mu+2M$ is the total mass. The Hamiltonian $H_{int}$ {\color{black}depends only on the relative
ccordinates and respective momenta and} describes the molecular dynamics with respect to
the reference frame at which the molecule center of mass (CM) is at rest. It
depends only on the internal variables,
since it commutes with the CM coordinate
$${\bf R}=\frac{1}{M_{total}}(\bfr_1+\bfr_2 \mu+ \bfr_3M+\bfr_4 M).$$

We have
$${\bf P}^2=-\sum_{i=1}^4\nabla_i^2-\sum_{i,j=1}^4\nabla_i\cdot\nabla_j.$$
To compute the expectation value of the CM energy, we need the following expressions
\begin{eqnarray*}&&\frac{\langle\Phi|(-\sum_{i=1}^4\nabla_i^2)|\Phi\rangle}{\langle\Phi|\Phi\rangle}
=\frac{I^{ee}_{KD}(\alpha,\gamma)+I^{ee}_{KD}(\gamma,\alpha)+2I^{ee}_{KE}(\alpha,\gamma)}{I^{ee}_{ND}(\alpha,\gamma)+I^{ee}_{NE}(\alpha,\gamma)}\\
&&\quad\quad\quad
+\frac{I^{PP}_{KD}(\beta,\beta')+I^{PP}_{KD}(\beta',\beta)+2I^{PP}_{KE}(\beta,\beta')}{I^{PP}_{ND}(\beta,\beta')+I^{PP}_{NE}(\beta,\beta')}
\\&&\frac{\langle\Phi|(-\sum_{i,j=1}^4\nabla_i\cdot\nabla_j)|\Phi\rangle}{\langle\Phi|\Phi\rangle}
=\frac{2I^{ee}_{p1p2E}(\alpha,\gamma)}{I^{ee}_{ND}(\alpha,\gamma)+I^{ee}_{NE}(\alpha,\gamma)}
+\frac{2I^{PP}_{p1p2E}(\beta,\beta')}{I^{PP}_{ND}(\beta,\beta')+I^{PP}_{NE}(\beta,\beta')},
\end{eqnarray*}
where $ I^{ee}_{p1p2E}(\alpha,\gamma)$ and  $ I^{PP}_{p1p2E}(\beta,\beta')$ are given in the Appendix.
Figure~\ref{fig5} illustrates the CM energy correction for the mono-muonated molecule, which is seen to fluctuate very little. Indeed, it remains almost constant.
\begin{figure}[h]
\centering
\includegraphics[width=.5\textwidth, height=0.3\textwidth]
{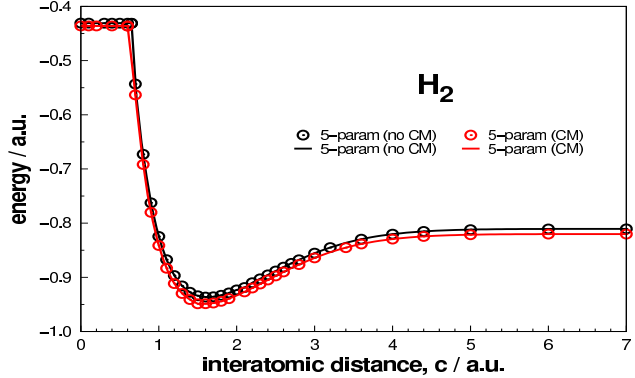}
\caption{
Upper curve: the energy of H$_2$ molecule, without CM corrections. Lower curve the energy of the H$_2$ molecule with CM correction.}
\label{fig7}
\end{figure}
The expectation value of $H$ is contaminated by the CM energy, while the expectation value of $H_{int}$ is not.
We notice that in the case of mono-muonic hydrogen, the contaminating contribution of the CM energy, although significant,
is almost independent  of $c$, {\color{black} as Figure~\ref{fig5} illustrates.}  We remark that, in the case of the  mono-muonic hydrogen, the minimum of the expectation value of $H$
 is about $-49.613\,\rm a.u.$ both for $c=0$ and for $c=0.3$. This means that, for $0<c<0.3$ the system is not properly described by any one of these configurations.
The appropriate description is then provided by a resonance between those configurations.
A similar observation applies to the expectation value of $H_{int}$, which is $-58.91\,\rm a.u.$ for $c=0$ and
 $-58.9\,\rm a.u.$ for $c=0.28\,\rm a.u.$ (see Figures 3 and 5). It seems very likely that the resonance effect will lead to a weak binding between the non-muonic hydrogen atom and the  muonic hydrogen atom, since the expectation values of $H$ and $H_{int}$ for $c=0$ almost coincide with the respective values for $c=\infty.$
{In the resonant state, the two protons are very close to each other, which is related to the catalytic property of the muon for fusion reactions.}
In the case of  non-muonic hydrogen, the contaminating contribution of the CM energy is extremely small; see Figure 6.
 On the other hand, in the case of  di-muonic hydrogen, the contaminating contribution of the CM energy fluctuates a lot with $c$,  (see Figures 2 and 7).
 Nevertheless, the binding energy is not strongly affected by the CM energy.
Indeed, the calculated binding energy is $7.63\,\rm a.u.$ if the CM correction is not taken into account and is $7.38\,\rm a.u.$ otherwise.

{\allowdisplaybreaks
\section{Outline}
It appeared timely to consider three four-particle molecules amongst the simplest that one may conceive to illustrate the meaning of the Born-Oppenheimer approximation at the occasion of its ninetieth aniversary.}
By considering molecular di-hydrogen and two of its relatives where one or both electrons are replaced by a negative muon, the radii of all particles relative to the center-of-mass around which they move immediately suggests that such an approximation is valid only in the case of the unsubstituted di-hydrogen molecule due to the large proton-to-electron mass ratio.
\begin{figure}[h]
\centering
\includegraphics[width=.5\textwidth, height=0.3\textwidth]
{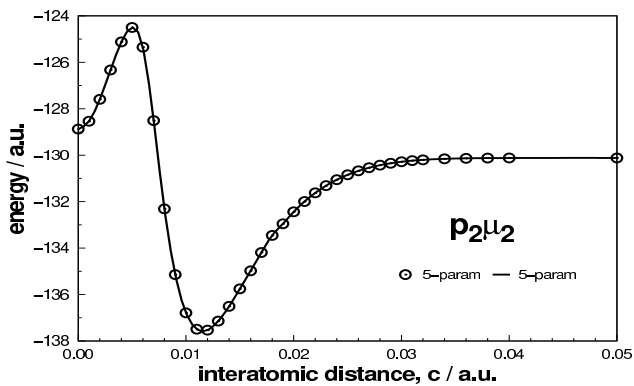}
\caption{
The energy of di-muonic molecule , with CM corrections. Minimization was performed after subtracting the CM energy. }
\label{fig8}
\end{figure}

\section*{Acknowledgments}
This work has the support of Funda\c{c}\~{a}o para a Ci\^{e}ncia e a Tecnologia, Portugal, via Coimbra Chemistry Centre through project PEst-OE/QUI/UI0313/2014
and via CFisUC 
under the projects No. UID/FIS/04564/2016 and POCI-01-0145-FEDER-029912.

\section*{Appendix}
The following integrals will be used
\begin{eqnarray*}
&&\color{black}I^{ee}_{VD}(\alpha,\gamma)
:=\int\int\d^3\r_1\d^3\r_2{\e^{-\alpha(\r_1-\frac{\bfc}{2})^2-\gamma(\r_2+\frac{\bfc}{2})^2}\over\sqrt{(\r_1-\r_2)^2}}=
{\pi^3\over(\alpha\gamma)^{3/2}c}{\rm erf}\left({\sqrt{\alpha\gamma}c\over\sqrt{\alpha+\gamma}}\right).\\
&&\color{black}I^{ee}_{VE}(\alpha,\gamma)=\int\int\d^3\r_1\d^3\r_2{\e^{-\frac{1}{2}(\alpha(\r_1-\frac{\bfc}{2})^2+\gamma(\r_1+\frac{\bfc}{2})^2)
-\frac{1}{2}(\alpha(\r_2-\frac{\bfc}{2})^2+\gamma(\r_2+\frac{\bfc}{2})^2)}\over\sqrt{(\r_1-\r_2)^2}}={8
 \pi^{5/2}\e^{-{\alpha\gamma c^2\over\alpha+\gamma}}\over(\alpha+\gamma)^{5/2}}
\\&&I^{ePD}_{VDD}(\alpha,\beta')=\int\int\d^3\r_1\d^3\r_2{\e^{-\alpha(\r_1-\frac{\bfc}{2})^2-\beta'(\r_2+\frac{\bfc}{2})^2}
\over\sqrt{(\r_1-\r_2)^2}}
={\pi^{3}\over\alpha^{3/2}\beta'^{3/2} c}{\rm erf}\left({\sqrt{\alpha\beta'} c\over\sqrt{\beta'+\alpha}}\right)\\
&&
I^{ePS}_{VDD}(\gamma,\beta')=\int\int\d^3\r_1\d^3\r_2{\e^{-\gamma(\r_1-\frac{\bfc}{2})^2-\beta'(\r_2-\frac{\bfc}{2})^2}
\over\sqrt{(\r_1-\r_2)^2}}
={2\pi^{5/2}\over\beta'\gamma\sqrt{\gamma+\beta'}}\\
&&I^{eP}_{VDE}(\alpha,\beta,\beta')=\int\int\d^3\r_1\d^3\r_2{\e^{-\alpha(\r_1-\frac{\bfc}{2})^2-{\beta\over2}
(\r_2+\frac{\bfc}{2})^2-{\beta'\over2}(\r_2-\frac{\bfc}{2})^2}
\over\sqrt{(\r_1-\r_2)^2}}\\
&&={4\pi^3\e^{-\beta\beta'\bfc^2\over2(\beta+\beta')}\over\alpha^{3/2}\beta\sqrt{2(\beta+\beta')}c}{\rm erf}\left(\beta\sqrt\alpha c\over\sqrt{(\beta+\beta')(2\alpha+\beta+\beta')}\right).\\
&&I^{eP}_{VEE}(\alpha,\gamma,\beta,\beta')=\int\int\d^3\r_1\d^3\r_2{\e^{-{\alpha\over2}(\r_1-\frac{\bfc}{2})^2-{\gamma\over2}(\r_1+\frac{\bfc}{2})^2
-{\beta\over2}(\r_2-\frac{\bfc}{2})^2-{\beta'\over2}(\r_2+\frac{\bfc}{2})^2}\over\sqrt{(\r_1-\r_2)^2}}
\\&&={8\pi^3\e^{-{\alpha\gamma c^2\over2(\alpha+\gamma)}-{\beta\beta'c^2\over2(\beta+\beta')}}
\over(\alpha\beta'-\gamma\beta)
\sqrt{(\beta+\beta')(\alpha+\gamma)}}{\rm erf}
\left({\alpha\beta'-\gamma\beta\over\sqrt{2(\alpha+\gamma)(\beta+\beta')(\beta+\beta'+\alpha+\gamma)}}\right)
\\
&&I_{KD}^{ee}(\alpha,\gamma)=\int\int\d^3\r_1\d^3\r_2(\mathbf{\nabla}_1\e^{-\frac{1}{2}\alpha(\r_1-\frac{\bfc}{2})^2})
\cdot(\mathbf{\nabla}_1\e^{-\frac{1}{2}\alpha(\r_1-\frac{\bfc}{2})^2})~\e^{-\alpha(\r_2+\frac{\bfc}{2})^2}\\
&&=\frac{3}{2}{\pi^3\alpha\over(\alpha\gamma)^{3/2}}
,\quad
I_{KD}^{ee}(\gamma,\alpha)
=\frac{3}{2}{\pi^3\gamma\over(\alpha\gamma)^{3/2}}
\\
&&I_{KE}^{ee}(\alpha,\gamma)=\int\int\d^3\r_1\d^3\r_2(\mathbf{\nabla}_1\e^{-\frac{1}{2}\alpha(\r_1-\frac{\bfc}{2})^2})
\cdot(\mathbf{\nabla}_1\e^{-\frac{1}{2}\gamma(\r_1+\frac{\bfc}{2})^2})
~\e^{-\frac{1}{2}(\alpha(\r_2-\frac{\bfc}{2})^2+\gamma(\r_2+\frac{\bfc}{2})^2)}\\
&&\color{black}
=\left({3\over(\alpha+\gamma)}-{\alpha\gamma\bfc^2\over(\alpha+\gamma)^2}\right)
{8\pi^3\alpha\gamma\over(\alpha+\gamma)^3}\e^{-{\alpha\gamma\bfc^2/(\alpha+\gamma)}}\\
&&I_{ND}^{ee}(\alpha,\gamma)
=\alpha^2\int\int\d^3\r_1\d^3\r_2\e^{-\alpha(\r_1-\frac{\bfc}{2})^2}
~\e^{-\gamma(\r_2+\frac{\bfc}{2})^2}
={\pi^3\over(\alpha\gamma)^{3/2}}
={\pi^3\over(\beta\beta')^{3/2}}
\\
&&I_{NE}^{ee}(\alpha,\gamma)
=\int\int\d^3\r_1\d^3\r_2
\e^{-\frac{1}{2}\alpha((\r_1-\frac{\bfc}{2})^2+\gamma(\r_1+\frac{\bfc}{2})^2+\alpha(\r_2-\frac{\bfc}{2})^2+\gamma(\r_2+\frac{\bfc}{2})^2)}
={8\pi^3\e^{-{\alpha\gamma c^2\over \alpha+\gamma}}\over(\alpha+\gamma)^3}
\\
&&I^{eP}_{NDD}(\alpha,\beta)=\int\int\d^3\r_1\d^3\r_2{\e^{-\alpha(\r_1-\frac{\bfc}{2})^2-\beta(\r_2+\frac{\bfc}{2})^2}}={\pi^3\over\alpha^{3/2}\beta^{3/2}}\\
&&I^{eP}_{NDE}(\alpha,\beta,\beta')=\int\int\d^3\r_1\d^3\r_2{\e^{-\alpha(\r_1-\frac{\bfc}{2})^2-{\beta\over2}(\r_2+\frac{\bfc}{2})^2
-{\beta'\over2}(\r_2-\frac{\bfc}{2})^2}
}={2^{3/2}\pi^3\e^{-\beta\beta'\bfc^2\over2(\beta+\beta')}\over(\alpha{(\beta+\beta')})^{3/2}}
\\
&&I^{eP}_{NEE}(\alpha,\gamma,\beta,\beta')=\int\int\d^3\r_1\d^3\r_2{\e^{-{\alpha\over2}(\r_1-\frac{\bfc}{2})^2-{\gamma\over2}(\r_1+\frac{\bfc}{2})^2
-{\beta\over2}(\r_2-\frac{\bfc}{2})^2-{\beta'\over2}(\r_2+\frac{\bfc}{2})^2}
}
={8\pi^3\e^{-{\alpha\gamma c^2\over2(\alpha+\gamma)}-{\beta\beta'c^2\over2(\beta+\beta')}}
\over
{((\beta+\beta')(\alpha+\gamma))^{(3/2)}}}\\
&&I^{ee}_{p1p2E}(\alpha,\gamma)=\int\int\d^3\bfr_1\d^3\bfr_2
\e^{-\frac{1}{2}(\alpha(\bfr_2-\bfc/2)^2+\gamma(\bfr_1+\bfc/2)^2)}
(-\nabla_1\cdot\nabla_2)
\e^{-\frac{1}{2}(\alpha(\bfr_1-\bfc/2)^2+\gamma(\bfr_2+\bfc/2)^2)}\\
&&=\frac{8\pi^3\gamma^2\alpha^2c^2}{(\gamma+\alpha)^5}\exp\left({-\frac{\gamma\alpha c^2}{\gamma+\alpha}}\right),\\
\end{eqnarray*}

The following functions are also needed
\begin{eqnarray*}
&&
I^{PP}_{VD}(x,y)=I^{ee}_{VD}(x,y),~~
I^{PP}_{KD}(x,y)=I^{ee}_{KD}(x,y),~~
I^{PP}_{VE}(x,y)=I^{ee}_{VE}(x,y),\\&&
I^{PP}_{KE}(x,y)=I^{ee}_{KE}(x,y),~~
I^{PP}_{NE}(x,y)=I^{ee}_{NE}(x,y),~~
I^{PP}_{p_1p_2}(x,y)=I^{ee}_{p_1p_2}(x,y),\\&&
I^{PP}_{ND}(x,y)=I^{ee}_{ND}(x,y),~~
I^{PP}_{NED}(x,y,z)=I^{ee}_{NDE}(x,z,y),~~
I^{PP}_{VED}(x,y,z)=I^{ee}_{VDE}(z,x,y),
\end{eqnarray*}
with the ones below being the same as the ones previously defined with other arguments,
\begin{eqnarray*}
I^{ePP}_{VDD}(\gamma,\beta),~~
I^{ePS}_{VDD}(\gamma,\beta'),~~
I^{eP}_{VDE}(\gamma,\beta',\beta),~~
I^{eP}_{VED}(\gamma,\alpha,\beta'),~~
I^{ee}_{KD}(\gamma,\alpha),~~
I^{PP}_{KD}(\beta',\beta).~~
\end{eqnarray*}
\color{black}


\end{document}